%% file: main.tex
\documentclass[runningheads]{llncs}
\usepackage[T1]{fontenc}
\usepackage{xcolor}
\usepackage{amsmath}
\usepackage{graphicx}
\usepackage{xspace}
\usepackage{algorithm, algpseudocode}
\usepackage{pifont}
\newcommand{\cmark}{\ding{51}}%
\newcommand{\xmark}{\ding{55}}%
\usepackage{subcaption}
\algtext*{EndIf}
\algtext*{EndFor}
\algtext*{EndProcedure}
\algrenewcommand\algorithmicthen{}

\newcommand{\basesys}{Chiron\xspace}
\newcommand{\sys}{\textsc{Anthemius}\xspace}
\newcommand{\sysmax}{240\%\xspace}

\begin{document}


\title{\sys: Efficient \& Modular Block Assembly for Concurrent Execution}

\author{Ray Neiheiser\inst{1} \and
Eleftherios Kokoris-Kogias\inst{2}}
%
\authorrunning{R. Neiheiser \and E. Kokoris-Kogias}
%
\institute{ISTA, Klosterneuburg, Austria \and
Mysten Labs, Athens, Greece}

\maketitle

\begin{abstract}

Many blockchains such as Ethereum execute all incoming transactions sequentially significantly limiting the potential throughput.
A common approach to scale execution is parallel execution engines that fully utilize modern multi-core architectures.
Parallel execution is then either done optimistically, by executing transactions in parallel and detecting conflicts on the fly, or guided, by requiring exhaustive client transaction hints and scheduling transactions accordingly.

However, recent studies have shown that the performance of parallel execution engines depends on the nature of the underlying workload.
In fact, in some cases, only a 60\% speed-up compared to sequential execution could be obtained. This is the case, as transactions that access the same resources must be executed sequentially. For example, if 10\% of the transactions in a block access the same resource, the execution cannot meaningfully scale beyond 10 cores. Therefore, a single popular application can bottleneck the execution and limit the potential throughput.

In this paper, we introduce \sys, a block construction algorithm that optimizes parallel transaction execution throughput. We evaluate \sys exhaustively under a range of workloads, and show that \sys enables the underlying parallel execution engine to process over twice as many transactions.

\end{abstract}

\input{1introduction.tex}

\input{2systemmodel.tex}

\input{3overview.tex}

\input{4evaluation.tex}

\input{6motivationandrelatedwork}

\input{5discussion.tex}

\input{7concl.tex}



\bibliographystyle{splncs04}
\bibliography{main}

\end{document}

%% file: 1introduction.tex
\section{Introduction}

The growing interest in blockchain and distributed ledger technology has resulted in many research advances in the field, ranging from improvements on the consensus layer~\cite{kauri,narwahl} to sharding~\cite{omniledger} and parallel transaction execution~\cite{blockstm,chiron}. 
As most blockchains still execute transactions sequentially, parallel smart contract execution engines that take advantage of modern multi-core architectures are considered a crucial building block to scale blockchain transaction throughput~\cite{blockstm}.

Existing approaches to parallel execution can be roughly divided into two categories: optimistic and guided. Optimistic approaches, such as Block-STM~\cite{blockstm}, are designed to execute transactions in parallel, detect conflicts as they arise, and re-execute affected transactions. 
However, in blockchain environments characterized by highly contended workloads~\cite{etheng,chiron}, conflicts arise more often, requiring more frequent re-executions of transactions.

In contrast to optimistic approaches, guided approaches strictly limit read/write access by requiring transactions to pre-declare an exhaustive list of resources (i.e., addresses) that will be accessed during execution. This allows the scheduler to identify independent transactions and execute them concurrently. Examples of this approach include FuelVM, Solana, or Sui~\cite{fuelvm,solana,sui}. While this avoids the re-execution overhead in settings with high contention, it puts additional load on the application developers. Furthermore, in some cases, it may not be possible to precisely predict at transaction creation time which resources will be accessed during execution, as the application state might change in the meantime. Then, an overly pessimistic approach is required, locking a wider range of resources, and potentially resulting in the sequential execution of transactions that otherwise could have been executed concurrently.

Combining both approaches, Polygon recently introduced an update~\cite{polygonupdate} that extracts transaction dependencies during block creation and includes this dependency tree as metadata in the block. This approach allows to optimize scheduling during the execution phase, avoiding unnecessary re-executions and pessimistic locking~\cite{polygonupdate}. Nonetheless, this approach requires executing transactions on the critical path of consensus during block creation, crippling the potential throughput.
A similar approach is \basesys~\cite{chiron} which leverages execution hints to speed up execution on struggling validators and full nodes. Chiron guarantees safety in the presence of invalid hints by utilizing the validation step of Block-STM, which identifies conflicting resource accesses and reschedules transactions that potentially accessed shared resources in parallel for re-execution~\cite{blockstm}.

However, as outlined in \cite{chiron}, due to the characteristics of blockchain workloads, transaction execution remains a significant bottleneck.
This is the case, as transactions that access the same resources must be executed sequentially and, as several recent studies have shown, in practice a significant portion of the transactions access the same resources, resulting in a long sequential path of transactions slowing down the system~\cite{ethparallelimpro,chiron}.
As such, the performance is currently limited by the workload.

Due to the nature of the problem, a single popular application can bottleneck the execution engine and cripple the throughput of the system~\cite{chiron}.
This could be a newly launched NFT, a popularly traded token, or even the on/off-boarding of a popular layer-2 smart contract.
This is further aggravated by the fact that most existing blockchains that support parallel execution currently have no pricing mechanisms to charge clients for accessing popular resources causing system bottlenecks.

Most blockchains such as Ethereum~\cite{ethereum} prevent extensive execution times by limiting the combined execution complexity in gas of each given block.
However, a single parameter is insufficient in the context of parallel execution, as it does not take transaction dependencies and potential parallelization into account. Therefore, a novel approach is necessary to make block assembly sensitive to transaction dependencies and execution complexity, charging clients for accessing popular resources and delaying transactions that would otherwise bottleneck the execution. 

In this paper, we propose \sys, a novel approach to construct blocks that takes both the execution complexity in gas and the distribution of resource accesses into account to construct ''Good Blocks'' that can be executed efficiently in parallel. We evaluate \sys extensively under a series of realistic workloads, showing a consistent speed-up up to \sysmax compared to native parallel execution.
\sys not only vastly improves the execution performance but also prevents popular or malicious applications from bottlenecking the system, eliminating a performance attack scenario.
\sys provides different latency paths between transactions accessing congested and not congested resources. Transactions can still be fast-tracked by paying higher transaction fees, resulting in a price that more closely reflects its resource consumption. We discuss this further in Section~\ref{sec:discussion}.

Moreover, thanks to its modular design, \sys can be integrated into any state-of-the-art blockchain seamlessly, without the need for a hard fork or modifications to the execution engine or consensus mechanism. \sys operates stateless and only requires execution hints such as the resources that will be accessed during execution. In blockchains such as Sui and Solana~\cite{sui,solana} these hints are already present during block construction, while in blockchains such as Aptos or Ethereum~\cite{aptos,ethereum} these hints could either be simulated in a pre-execution step or generated at the full nodes.

In summary, we provide the following contributions:

\begin{itemize}
    \item We propose \sys, a novel and modular block construction algorithm and approach to speed up parallel execution without security tradeoffs.
    \item We evaluate \sys integrated with both an optimistic and a guided execution engine under the \basesys benchmarks resulting in a significant speed-up in almost all settings.
\end{itemize}

In Section~\ref{sec:systemmodel}, we present the System Model of \sys, followed by a detailed overview of \sys in Section~\ref{sec:overview}. Next, we describe the implementation and evaluation in Section~\ref{sec:evaluation}. Related work is reviewed in Section~\ref{sec:relatedwork}, and potential drawbacks, along with their solutions, are discussed in Section~\ref{sec:discussion}. Finally, we conclude the paper in Section~\ref{sec:conclusion}.

%% file: 2systemmodel.tex
\section{System Model}
\label{sec:systemmodel}

We assume a blockchain environment consisting of $N$ server processes $p_1, p_2,..,p_N$ and $I$ client processes $c_1, c_2,..,c_I$. Clients send signed transactions to the server processes to be included in a future block.
The blockchain functions as the Public Key Infrastructure where the identifier of a client is its public key, and clients use their private keys to sign their transactions.

We assume a consensus abstraction as a blackbox, where one or more processes construct blocks of transactions and propose them to the consensus mechanism. As a result, the consensus abstraction outputs an ordered sequence of blocks $b_1, b_2, \dots, b_n$, which is then processed by the execution engine. Additionally, we assume an execution engine abstraction as a blackbox that receives this ordered sequence of blocks from the consensus abstraction and executes them deterministically.

In the context of this work, we make no assumptions regarding the coupling between the consensus and execution layers. The interaction between consensus and execution may either follow a modular, decoupled approach, as in Sui and Aptos~\cite{sui,aptos}, or operate in a tightly coupled, sequential manner, as in Ethereum~\cite{ethereum}.

Client transactions might range from simple peer-to-peer transactions to complex application logic with the help of smart contracts. As applications might access arbitrary resources (i.e., addresses) that can not easily be deduced, we assume the existence of a system that provides hints about the resources a transaction will access during execution to the block producer. This can either be in the form of client hints as in Solana or Sui~\cite{solana,sui}, or in the form of an optimistic pre-execution step that determines these hints locally as in Polygon~\cite{polygonupdate}. However, we do not assume the list of hints to be exhaustive or correct. Transactions with incomplete or incorrect hints might trigger re-executions if the execution engine is Block-STM or a derivative~\cite{blockstm,chiron}, or aborted in Solana or Sui~\cite{solana,sui}.

%% file: 3overview.tex
\section{Anthemius}
\label{sec:overview}

The primary objective of \sys is to redesign the block-assembly approach in blockchains that offer parallel execution to improve the overall system throughput and prevent popular applications from creating bottlenecks by factoring in transaction dependencies and execution time.

At the time of writing, most blockchains that support parallel transaction execution use a single parameter such as the computational complexity in gas, the raw block size in bytes, or the number of transactions to limit the block size~\cite{aptos,sui,solana}. However, in the context of parallel transaction execution, a single parameter does not reflect the execution complexity of a block. If all transactions in the block access the same resource, the execution time is the sum of the runtime of all transactions. In contrast, if none of the transactions access conflicting resources, the runtime depends on the number of cores.

Therefore, as a first step to begin constructing ``Good Blocks'', we need parameters that allow us to quantify this. We deploy two parameters to address this. First a transaction complexity parameter in Gas, similar to Ethereum, and second a concurrency parameter $c$ describing the system's ability to execute transactions in parallel (i.e. number of cores). As a result, the total maximum capacity of each block is $c * Gas$.

In the next sections, we first discuss where \sys fits into existing blockchain architectures. Following that, we outline the design of the block construction algorithm that considers both parameters and constructs blocks sensitive to transaction dependencies and their execution time to speed up the parallel execution of the block.

\subsection{Architecture}

\begin{figure}
\begin{center}
	\includegraphics[width=1\columnwidth]{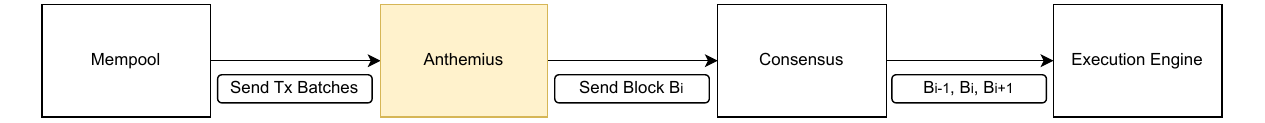}
\end{center}
\caption{\sys is inserted between the Mempool and Consensus}
\label{fig:architecture}
\end{figure}

Figure~\ref{fig:architecture} shows where \sys fits into the existing protocol stack of a blockchain. \sys is a modular layer that can be inserted between the consensus layer and the mempool where client transactions are stored and handled. 
In \sys, instead of fetching transactions directly from the mempool, the consensus layer fetches blocks of transactions through \sys. In turn, \sys obtains its transactions from the mempool, divides transactions into batches, and constructs the block to return to consensus.
Following that, the block is proposed in consensus which outputs an ordered list of blocks to the execution engine.

\sys requires the read and write sets of transactions, as well as an estimation of their execution time, to assess dependencies between transactions and construct blocks that can be executed efficiently in parallel. This information is already available in blockchains such as Solana~\cite{solana} and Sui~\cite{sui}, where transactions must declare all resource addresses they access during execution. In other blockchains, such as Ethereum~\cite{ethereum}, this information can be obtained, for example, by executing the transactions.

This design allows \sys to be seamlessly integrated into any existing blockchain stack with minimal architectural and system changes, and without changing the block structure. Furthermore, since \sys operates solely on the set of transactions, their read and write sets, and their gas footprints, it remains essentially stateless. This makes \sys particularly suitable for deployment in modular architectures, such as Narwhal, where only the execution layer is stateful~\cite{narwahl}.  

\subsection{Block Construction}

An important problem that has to be tackled when constructing good blocks is the absence of information regarding the structure of the current workload. If all transactions in the mempool access the same resources, attempting to schedule them efficiently can further slow down an already bottlenecked system. Similarly, if the algorithm is too strict in situations where a large percentage of transactions access the same resources, the synergetic effects of executing larger batches of transactions are lost. This is the case, as, for each block, the system has to instantiate the executor and worker threads, set up the virtual machine, extract the execution results, etc.

Therefore, as a first step, we divide \sys into two modular elements. First, the \textit{batch handler}, which polls batches of transactions from the mempool and hands the batches to the \textit{batch scheduler} in a batch-by-batch fashion. Second, the \textit{batch scheduler}, that attempts to include a given batch into the current block and provides feedback to the \textit{batch handler} about the success rate.
Subsequently, based on the feedback, the batch handler can adjust the inclusion policy to prevent too small blocks and also avoid wasting scheduling time on difficult-to-schedule workloads.

\paragraph{Batch handler.}

\begin{algorithm}[t]

\centering
\caption{\textsl{Batch Handler}}
\label{algo:batchhandler}
\begin{algorithmic}[1]

\Procedure{CreateGoodBlock}{$block, maxgas, c$}
\State{$seqlimit = \frac{maxgas}{c}$} \Comment{Limit on the sequential path}
\State{$resmap \gets \emptyset$} \Comment{Map to track transaction dependencies}
\State{$skippedclients \gets \emptyset$} \Comment{Set to track clients with skipped transactions}
\State{$numrelax \gets 0$} \Comment{Number of times inclusion rate was relaxed}
\For{\textbf{all} $\textit{batch} \in mempool$}
\State{$incrate \gets$ \textsc{schedule($block, batch, seqlimit, c, resmap, skippedclients$)}} \label{algo:batchhandler:schedule}
\If{$incrate < \textsc{targetincrate} $}
\If{$numrelax \geq \textsc{maxrelaxnum} \lor (incrate = 0 \land batch.isfull)$} \label{algo:batchhandler:check}
\State{\textbf{return}}
\EndIf
\State{$seqlimit = \frac{maxgas}{c} * \textsc{min}(\textsc{maxrelaxrate}, incrate*\textsc{targetincrate})$} \label{algo:batchhandler:relax}
\State{$numrelax++$}
\EndIf
\EndFor

\EndProcedure
\end{algorithmic}
\end{algorithm}

The functionality of the Batch Handler is outlined in Algorithm~\ref{algo:batchhandler}. The batch handler receives a $block$ to fill, the global concurrency parameter $c$, and the maximum gas limit. It then calculates a limit on the sequential path $seqlimit$ and initiates a map to track the transaction dependencies $resmap$ as well as a set of clients with skipped transactions $skippedclients$. 

Next, the batch handler retrieves transaction batches from the mempool and hands them to the batch scheduler alongside the block, the limit on the gas, the number of cores, the transaction resource dependencies $resmap$, and $skippedclients$ set in Line~\ref{algo:batchhandler:schedule}.
The batch scheduler responds with the transaction inclusion rate $incrate$.

Depending on the workload, as mentioned, the $seqlimit$ may be very strict which can result in very few transactions being included in a block.
Therefore, if the inclusion rate $incrate$ is smaller than some $\textsc{targetincrate}$, we relax the gas limit relative to the inclusion rate, up to some \textsc{maxrelaxrate} (Line~\ref{algo:batchhandler:relax}).

However, if the inclusion rate was too small for several consecutive attempts (i.e. $numrelax \geq \textsc{maxrelaxnum}$), we exit scheduling to avoid building a heavily sequential block again.
Furthermore, if there was an attempt to schedule a full batch and no transaction of this batch was successfully included in the current block ($incrate = 0$) we also stop scheduling (Line~\ref{algo:batchhandler:check}) as this indicates that at this point transactions are only included at a high cost to execution performance and scheduling latency. The rest of the transactions are then only included in a later block.


\paragraph{Batch Scheduler.}

Scheduling transactions with interdependencies and varying runtimes is a known NP-complete problem~\cite{BAKER1996225} where approximate solutions can construct near optimal schedules in polynomial time. However, polynomial runtime, particularly when executed on the critical path of consensus, may lead to a construction time that outweighs the performance gains achieved from producing ''Good Blocks.''

Fortunately, our first insight is that a near-optimal schedule for block construction is unnecessary. Instead, our main objective is to prevent popular resources and applications from creating a bottleneck while maximizing the parallel execution. We can achieve this by iterating over the set of resources each transaction accesses, recording the cost of the sequential path leading up to the transaction, and deciding if the transaction should be included in the current block by comparing the cost of the path with the gas per core parameter.
Furthermore, we also want to delay transactions that access multiple hot resources as they make it harder to schedule subsequent transactions.

As a result, the complexity of the block construction is of $O(N*k)$ where $N$ is the number of transactions and $k$ is the average number of resource accesses per transaction.

\begin{algorithm}[t]

\centering

\caption{\textsl{Batch Scheduler - Called in Line~\ref{algo:batchhandler:schedule} of Algorithm~\ref{algo:batchhandler}}}
\label{algo:batchscheduler}
\begin{algorithmic}[1]

\Procedure{schedule}{$block, batch, seqlimit, c, resmap, skippedclients$}
\For{\textbf{all} $\textit{tx} \in batch$} \Comment{Iterate over transactions} \label{algo:batchscheduler:iterate}
     \If{$tx.sender$ \textbf{in} $skippedclients$}
            \State{\textbf{continue}} \Comment{Skip transaction inclusion} \label{algo:batchscheduler:skipuser}
    \EndIf
    \State{$chaincost \gets 0$} \Comment{Longest chain length}
    \State{$hotresources \gets 0$}
    \For{\textbf{all} $\textit{readres} \in tx.readset$} \Comment{Iterate over readset} \label{algo:batchscheduler:readloop}
        \If{$readres \in resmap$} \Comment{Find longest chain}
            \If{$resmap[readres] > chaincost$} \Comment{Find read with largest cost}
                \State{$chaincost \gets resmap[readres]$}
            \EndIf
            \If{$resmap[readres] > \frac{block.gas}{c}$} \Comment{Check if read exceeds limit}
                \State{$hotresources++$} \label{algo:batchscheduler:hotread}
            \EndIf
        \EndIf
    \EndFor
    \If{$hotresources \geq \textsc{maxhotr} \land (|block| > \textsc{lim} \lor |block| < \textsc{maxlen}-\textsc{lim})$}  \label{algo:batchscheduler:hotreadcheck}
        \State{$skippedclients \gets skippedclients \cup tx.sender$}
        \State{\textbf{continue}} \Comment{Skip transaction inclusion}
    \EndIf
    
    \If{$chaincost + tx.gas > seqlimit \lor block.gas + tx.gas > seqlimit*c$}  \label{algo:batchscheduler:gascheck}
            \State{$skippedclients \gets skippedclients \cup tx.sender$}
            \State{\textbf{continue}} \Comment{Skip transaction inclusion}
    \EndIf
    \State{$block \gets block \cup tx$} \Comment{Add tx to Block}
    \For{\textbf{all} $\textit{writeres} \in tx.writeset$} \Comment{Iterate over writeset}
        \If{$writeres \notin resmap \lor resmap[writeres] < chaincost$}
         \State{$resmap[writeres] \gets chaincost$} \Comment{Note new chain length}
        \EndIf
    \EndFor
\EndFor
\State{\textbf{return}$(\frac{numscheduled}{|batch|})$}
\EndProcedure
\end{algorithmic}
\end{algorithm}

Algorithm~\ref{algo:batchscheduler} shows how we achieve this. The algorithm starts with the call of the \textsc{schedule} method, which receives the block to include the transactions in, the batch of transactions to schedule, the maximum gas per core $seqlimit$, the concurrency parameter $c$, the map of resources and the skipped clients. Following that, it starts iterating over all transactions in the batch (line~\ref{algo:batchscheduler:iterate}).
First, to maintain the order clients specified (e.g. through sequence numbers), after a client had a transaction skipped, the client is added to the $skippedclients$ set and no further transaction from this client will be included in this block.(Line~\ref{algo:batchscheduler:skipuser}).
Following that, we iterate over all reads in the transaction read-set and attempt to calculate the read with the longest path in gas leading up to this transaction (Line~\ref{algo:batchscheduler:readloop}).
In parallel, we count the number of \textit{hot} reads. A hot read is a read on a resource that is accessed significantly more often than other resources. 

After this, we check whether the number of hot reads exceeds a predefined threshold, \textsc{maxhotr}. If this condition is met and the transaction is not within the first or last \textsc{lim}(i.e. 10\%) transactions, we skip the transaction (Line~\ref{algo:batchscheduler:hotreadcheck}).
We delay transactions with too many hot reads as they unify several critical paths of transactions which can severely bottleneck the execution.
However, we initially allow any transactions to be included up to some threshold \textsc{lim} to accumulate sufficient data to assess the complexity of reads and to guarantee that transactions that access several hot resources are eventually included. Furthermore, we also allow including transactions with multiple hot reads towards the end of the block as the block is almost full already and they are less likely to cause scheduling problems at this point.

Following that, we check if the transaction cost itself is larger than the max gas per core $seqlimit$ or if the current transaction exceeds the total gas limit of the block. If so, we also skip the transaction (Line~\ref{algo:batchscheduler:gascheck}).

Finally, we include the transaction in the block, iterate over its write set, and record the transaction path cost in the resource map $resmap$ if its writes increase the critical path.
This results in an algorithm that is linear to the number of transactions per block, as the map accesses are $O(1)$ and we check each transaction at most once per block.

%% file: 4evaluation.tex
\section{Evaluation}
\label{sec:evaluation}


We implemented \sys on top of Block-STM~\cite{blockstm} and \basesys~\cite{chiron} in Rust to evaluate its performance impact on both an optimistic execution engine and a guided execution engine, covering two of the most widely adopted approaches to parallel execution in the blockchain space.
The implementation is publicly available on Github~\footnote{https://github.com/ISTA-SPiDerS/Anthemius}.
As \basesys is built on top of Block-STM, this simplifies the implementation and allows for an easier comparison of the results.
Furthermore, we use the parallel execution benchmarks proposed in \basesys~\cite{chiron}. 



Finally, we implemented the batch handler ($\sim70$ lines of code) and the batch scheduler ($\sim120$ lines of code) to assemble blocks and then forward these blocks to the respective execution engines.




\subsection{Benchmark}

The experiments were executed on a Debian GNU/Linux 12 server with two AMD EPYC 7763 64-Core Processors and 1024 GB of RAM. We generated batches of transactions with different distributions of read/write-accesses and different user distributions with the help of \basesys~\cite{chiron} for all five proposed workloads. Namely, one peer-to-peer workload (P2PTX), two Decentralized Exchange Workloads (DEXAVG and DEXBURSTY), one NFT workload (NFT), and one mixed workload (MIXED). 
These workloads are derived from real-world data from Ethereum and Solana and are designed to evaluate parallel transaction execution engines under realistic levels of contention. 
Each workload has a unique and realistic resource access pattern, along with a varying count of read and write operations per transaction.

Each experiment was executed a total of 10 times and the results we outline in this section present the average of all 10 runs. Furthermore, in each workload, we vary the number of worker threads from 4 to 32 in increments of 4.
Finally, we are interested in two key metrics: throughput, to assess the performance improvement introduced by \sys, and latency, to determine the average delay introduced by \sys.

We set the following parameters for the batch handler and batch scheduler:
First, we evaluate the execution engines using blocks of up to $\textsc{maxlen} = 10{,}000$ transactions, as this block size represents a sweet spot for both engines, where the execution setup overhead (e.g., virtual machine initialization) becomes negligible. Accordingly, we configured the batch size to match the target block size, as smaller batch sizes increase block construction overhead, while larger batch sizes reduce the batch handler's flexibility to adapt to the workload's characteristics.

Next, to minimize tail latency for transactions accessing hot resources, we allow the first and last $\textsc{lim} = 1{,}000$ transactions to be included freely without restrictions.  
Furthermore, we permit up to $\textsc{maxrelaxnum} = 2$ relaxations of the inclusion rate as we observed diminishing returns from additional relaxations and large scheduling costs beyond this point.  
We set the relaxation rate to a maximum of $\textsc{maxrelaxrate} = 100$, targeting an inclusion rate of $\textsc{targetincrate} = 2\frac{\text{maxlen}}{c}$. This accounts for the higher returns from a more aggressive target inclusion rate as the concurrency potential increases.  
Finally, we configure $\textsc{maxhotr} = 4$ to avoid uniting too many critical paths of transactions, ensuring manageable contention levels.




\subsection{Throughput}

As \sys delays the inclusion of some transactions in favor of others to enhance system performance, we provide the batch handler with several batches of $10{,}000$ transactions to saturate the system and measure the maximum throughput. Each batch is generated with the same distribution of resource accesses, both within and across batches. We then evaluate \sys by passing all batches to the batch handler and run \sys until all transactions from the first batch are successfully executed. Consequently, the evaluation for \sys spans multiple blocks, where the reported throughput represents the average throughput over the entire runtime and accounts for scheduling and execution time.
For the baseline versions of Block-STM and \basesys, we use a single block containing $10{,}000$ transactions that also fully saturates the system, with runtime variations dependent solely on the specific workload.

As blockchains such as Aptos or Sui decouple consensus from execution, block scheduling could be moved outside of the critical path of consensus. This can significantly reduce the overhead, as scheduling requires only a single thread and only has to be done at the proposer node. Due to this, we display two lines for \sys. First, one that serves as a ceiling on performance, where we assume that there is an idle thread that can be used for scheduling outside of the critical path of consensus, denoted \textit{Decoupled \sys}. Second, one that serves as a floor on performance where we count the full scheduling overhead on the critical path of consensus, referred to as \textit{\sys}.

\begin{figure*}[t]
\begin{subfigure}{0.5\textwidth}
\includegraphics[width=1\linewidth]{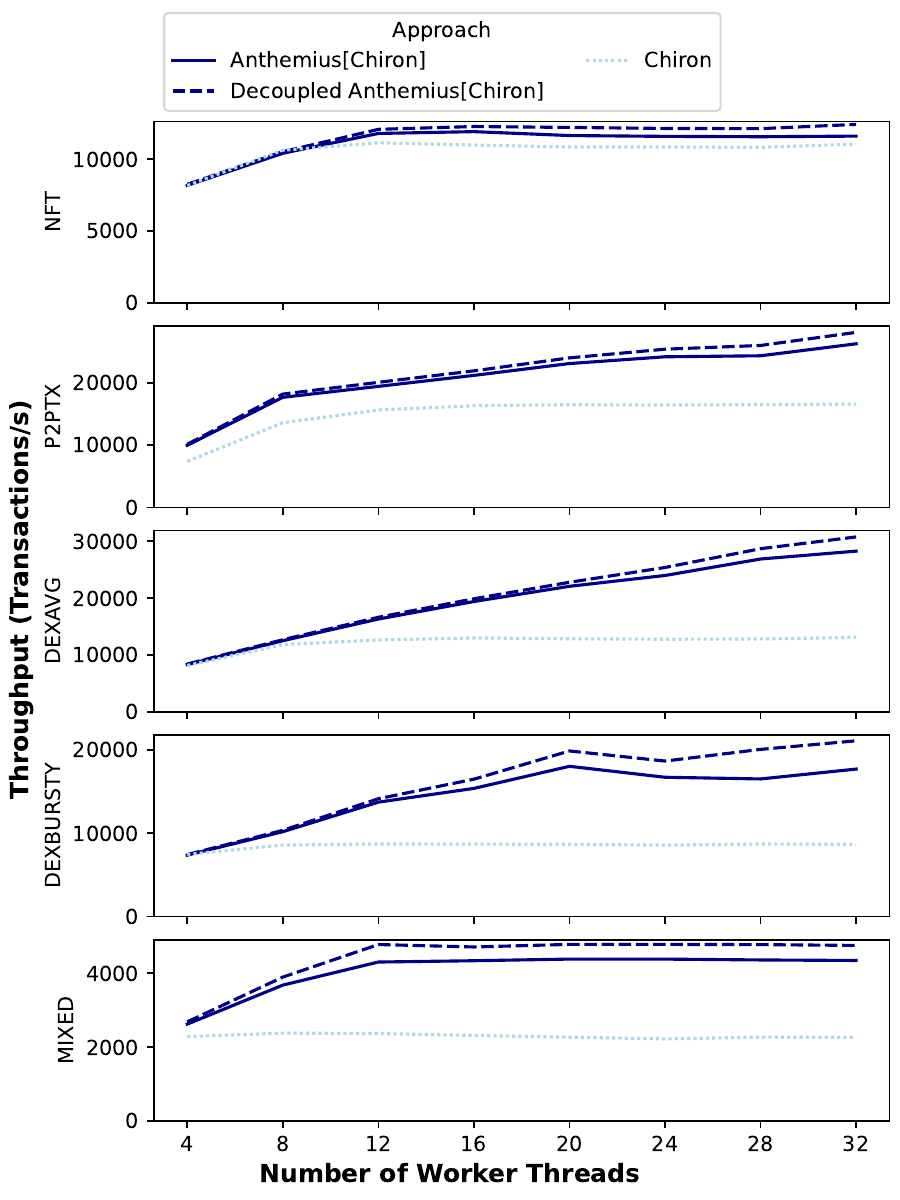} 
\caption{Throughput per Second - \basesys}
\label{fig:tputpythia}
\end{subfigure}
\begin{subfigure}{0.5\textwidth}
\includegraphics[width=1\linewidth]{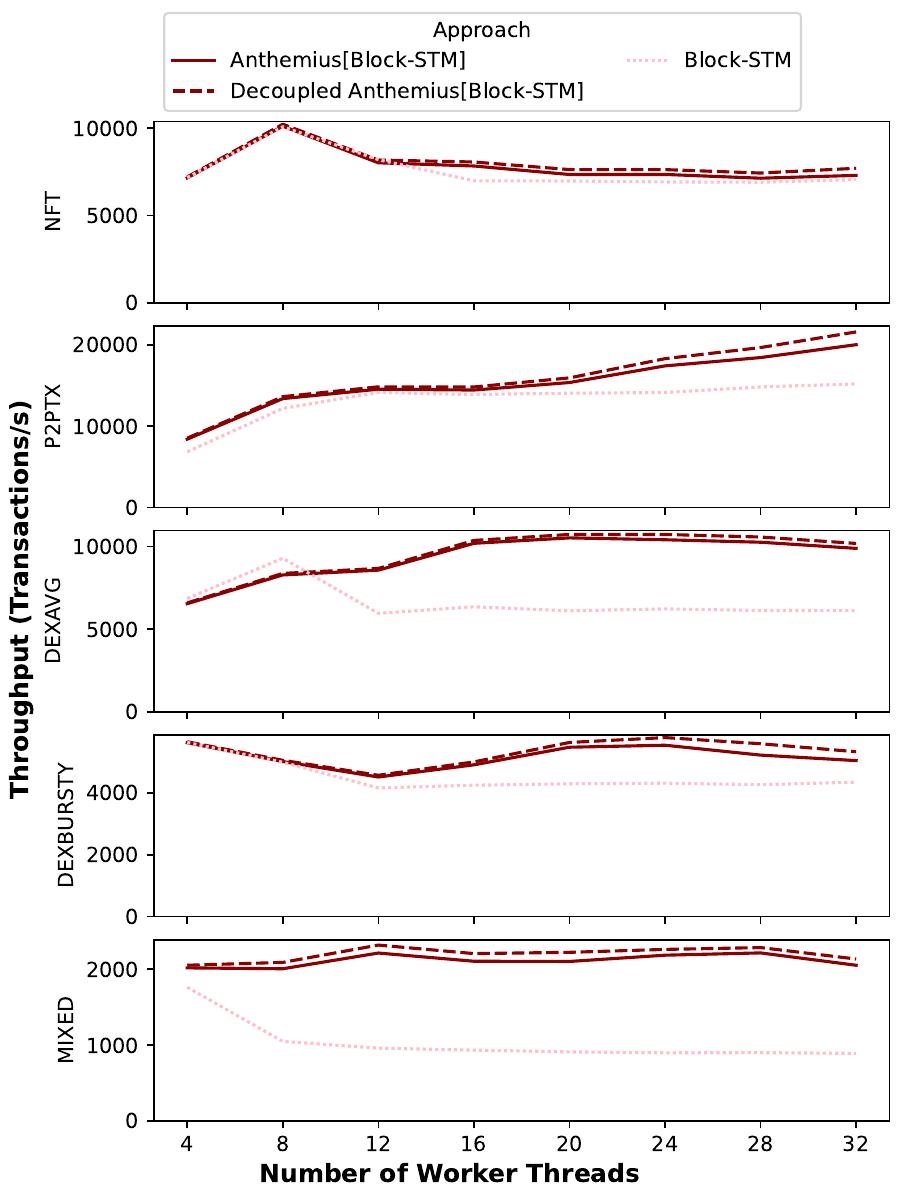}
\caption{Throughput per Second - Block-STM}
\label{fig:tputblockstm}
\end{subfigure}
\caption{Throughput per Second}
\label{fig:image2}
\end{figure*}

The results for \sys with \basesys are shown in Figure~\ref{fig:tputpythia}, with the throughput in transactions per second on the y-axis and the number of worker threads on the x-axis. With the NFT workload, we only see a small speedup from creating good blocks. This is due to the account distribution in this workload, where transactions from users appear very frequently in several batches. Due to this, once a transaction of a given user is skipped, the following transactions also have to be skipped, resulting in long scheduling times and leaving very few transactions behind that can be included in the block. In comparison, in the peer-to-peer workload there is already a significant improvement, where with an increasing number of worker threads, we can reach almost twice the initial throughput. Following that, with increasing contention and less repetitive users, the decentralized exchange workloads reach over 240\% performance boost compared to vanilla \basesys. While in the average DEX workload, the scheduling overhead is very small, with increasing contention and increasing number of worker threads we can also see an increased scheduling overhead.
Finally, in the mixed workload, we also see a large performance advantage. This is also due to the much higher overall execution complexity compared to the scheduling overhead. Due to the complexity of the workload, the overhead is constant after 12 cores, but \sys under this workload shows over 200\% performance advantage compared to vanilla \basesys.

The throughput results for \sys with Block-STM are shown in Figure~\ref{fig:tputblockstm}, with the throughput in transactions per second on the y-axis and the number of worker threads on the x-axis. 
Compared to the results with \basesys, the results for Block-STM vary more as the high contention within each block results in a large re-execution overhead. 
As such, even when we build better blocks with \sys, the contention in the block is still so high, that Block-STM struggles to take advantage of that.
We can still see the largest disadvantage in the NFT workload, due to the user distribution preventing us from building better blocks. Furthermore, we can see that in the peer-to-peer workload, once we reach 20 threads, \sys is starting to be able to compensate for the re-execution overhead of Block-STM and reach a speed-up of up to 25\%. Similarly, for the DEX workloads, there is an initial performance drop due to the re-execution overhead, which is only compensated with more worker threads later.
Finally, in the MIXED workload, \sys shows a constant speed up compared to vanilla Block-STM up to 200\% the original performance.


\subsection{Latency}

\begin{figure}[t]
\includegraphics[width=1\linewidth]{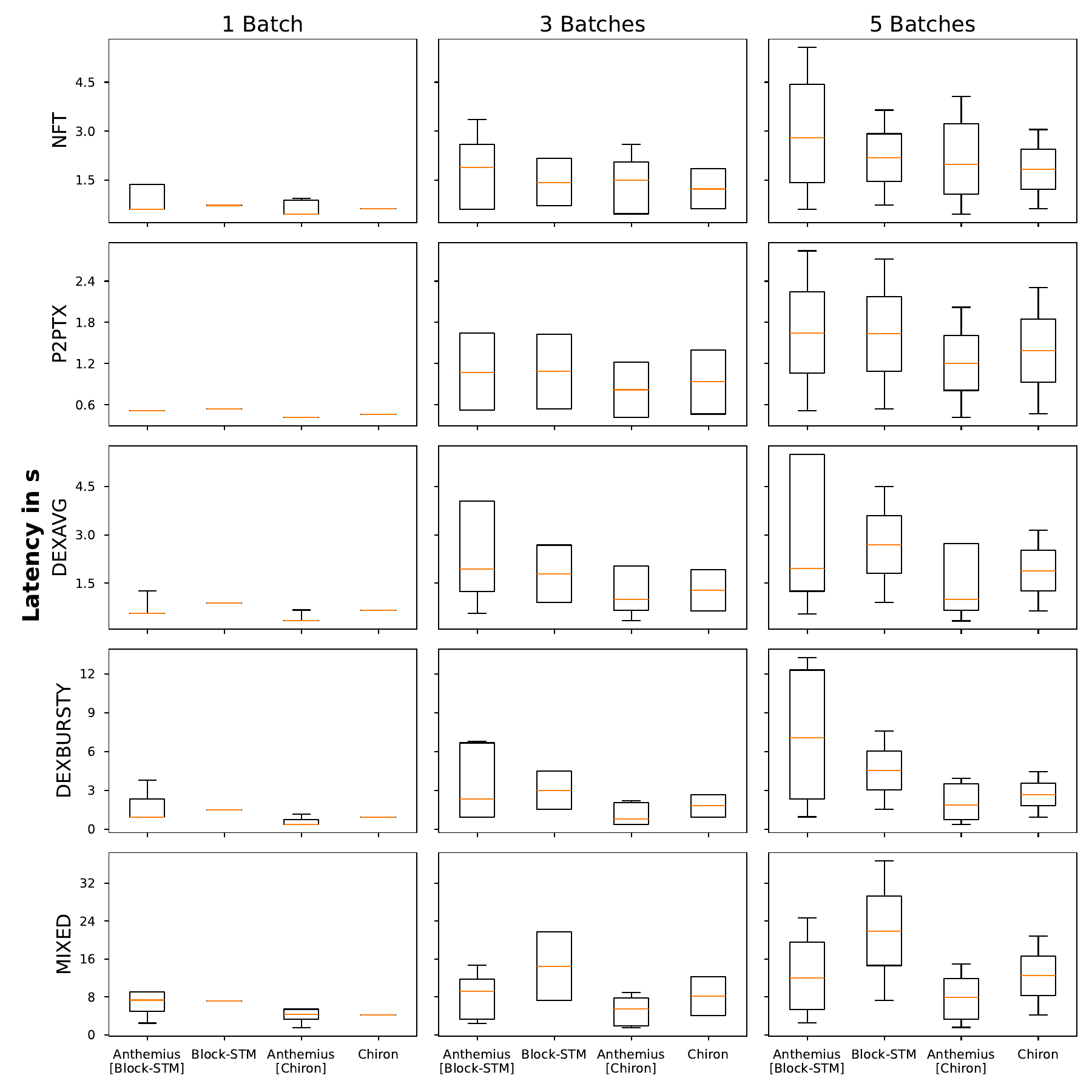} 
\caption{Tail Latency for \basesys and Block-STM}
\label{fig:latency}
\end{figure}

As we are delaying the inclusion of some transactions that access hot resources, we expect a latency overhead increase at the tail. 
Similarly to the throughput evaluation, we send several batches of transactions to the batch handler. To fully assess the effect of \sys, we evaluate how the tail latency develops when awaiting the finished execution of up to five batches for all workloads with a fixed number of $16$ cores.
The results of this evaluation are shown in Figure~\ref{fig:latency}, where the yellow line indicates the 50th percentile (median), the box represents the 25th and 75th percentiles (interquartile range), and the whiskers denote the 10th and 90th percentiles.

The results mirror what we saw in the throughput evaluation where in almost all workloads and configurations where \sys shows a significant speedup the average transaction latency is significantly lower. Furthermore, thanks to the large throughput advantage in these settings, especially when paired with \textit{Chiron}, \sys has a latency advantage for up to the 90\% percentile of transactions.

On the other hand, as expected, \sys shows a growing tail latency with an increasing number of batches. This
is expected since the congestion caused by the highly contended workloads results in different scheduling decisions. Nevertheless, we can see that the growing tail latency affects not only \sys but also the reference systems, although for certain workloads the effects of \sys are more prominent at the p90 percentile. 

This is a tradeoff the blockchain needs to take into account based on their expected workload and tune \sys parameters to better match the chracteristics of the transactions expected.

\subsection{Summary}

In this section, we evaluated the throughput improvement \sys can provide across different execution engines. Our findings demonstrate that while \sys improves throughput for both types of execution engines under several of the workloads, its impact is significantly larger when combined with guided execution engines. In this case, \sys provides a large throughput improvement across all but one of the workloads. The only exception is the NFT workload, where many high-frequency users appear across multiple blocks, preventing \sys from effectively rescheduling their transactions.

When it comes to latency, we analyzed the tail latency percentiles of delayed transactions. Our results show that for most workloads the majority of transactions (over 75\%) have lower or similar latency compared to the vanilla execution, while only the slowest 25\% of transactions sustain a latency overhead.
This indicates that \sys can be a valuable addition to any blockchain with a parallel execution engine where the workload does not primarily stem from a very small set of users.

%% file: 6motivationandrelatedwork.tex
\section{Related Work}
\label{sec:relatedwork}

To the best of our knowledge, there is no academic work proposing algorithms to construct blocks sensitive to parallel execution efficiency. 
While the problem is an NP-Complete scheduling problem which is explored in theoretical computer science~\cite{BAKER1996225}, the greedy version of these algorithms still requires polynomial time which would present a large overhead and negate most of the positive effects.
By relaxing the optimality requirement, instead, \sys achieves a linear complexity relative to the number of transactions per block.

In the database literature, there are numerous approaches to re-order transactions for reduced abort rates.
Most of the work in this context reorders transactions after execution to increase the goodput. Examples of this approach are Aria~\cite{aria}, where an efficient algorithm reorders transactions after execution based on the read and write sets to reduce the number of aborted transactions. Similarly, Sharma et al.~\cite{blurring} focus on execute-order blockchains where transactions are reordered during block construction.
While these approaches are efficient and can increase the goodput, none of them consider the parallel execution setting.

Eve~\cite{eve} is the most similar approach to \sys. In Eve, transactions are organized into batches such that, with high probability, no two transactions within the same batch access the same resource. This allows the execution engine to execute the block concurrently without having to worry about concurrent accesses during execution. Although the scheduling is very efficient, this approach is unsuitable for blockchain ecosystems where we are expecting a large percentage of transactions to overlap~\cite{chiron} and already have execution engines that can process transactions with dependencies efficiently.

\begin{table}[t]
\centering
{
\caption{Comparison of existing Block Production Approaches.}
\label{tab:comparison}
\begin{tabular}{|l|c|c|c|c|}
\hline
\textbf{Approaches} &  \textbf{Parallel} &  \textbf{Two-Dimensional} & \textbf{Dependency} & \textbf{Execution-Time} \\
 
 & \textbf{Execution} & \textbf{Gas Parameter} & \textbf{Sensitive}  & \textbf{Aware} \\

 \hline
  Ethereum~\cite{ethereum}  & \xmark  & \xmark  & \xmark & \cmark  \\ 
  Polygon~\cite{polygonupdate}  & \cmark  & \xmark  & \xmark & \cmark \\ 
   Aptos~\cite{aptos}  & \cmark  & \xmark  & \xmark & \xmark  \\ 
   Solana~\cite{solana} & \cmark   & \xmark   & \xmark & \cmark  \\ 
  \sys~ & \cmark   & \cmark   & \cmark& \cmark \\ 
\hline
\end{tabular}
}
\end{table}

We, therefore, focus on the current state of block assembly in production blockchains. The discussion is summarized in Table~\ref{tab:comparison}.
While Ethereum~\cite{ethereum} does not natively support parallel execution at this moment, it constructs its blocks sensitive to the execution complexity of the smart contracts. The version of Polygon~\cite{polygonupdate} with Block-STM integration supports parallel execution and takes the execution complexity into account. However, it only has a one-dimensional gas parameter and does not take dependencies into account. Aptos~\cite{aptos} supports parallel execution but is unaware of the execution complexity of the transactions at block construction time and only takes the number of transactions and byte size into account. In comparison, Finally, Solana~\cite{solana} also offers parallel execution and takes the execution complexity into account. However, Solana does not take dependencies into account and only limits the combined computational complexity of transactions of a given client. 


Therefore, to the best of our knowledge, \sys is the first work proposing a modular and practical algorithm for ``Good Block'' construction in the context of parallel smart contract execution.

%% file: 5discussion.tex
\section{Discussion}
\label{sec:discussion}

While \sys can achieve a performance boost of over \sysmax, there are tradeoffs. In this section, we discuss these tradeoffs and potential solutions.

\subsection{Malicious Leader}

While a correct leader can construct blocks that significantly speed up the system, the opposite is true for malicious leaders. In \sys the leader constructs the block sensitive to the number of cores and a gas per core measure, however, no mechanism in \sys enforces the leader to construct a block following this blueprint. Even though this might seem like an oversight, it is impossible to distinguish between a correct leader handling a fully sequential workload and a malicious leader deliberately constructing a sequential block.

This issue is inherent to blockchains that support parallel execution, such as Solana, Sui, or Aptos~\cite{solana,sui,aptos}. Two potential approaches could mitigate this challenge. One approach involves an expensive combination of a fair ordering protocol~\cite{fairordering} and a pre-execution stage on the critical path of consensus such as Pompe~\cite{zhang2020byzantine}. Alternatively, leaders could be incentivized through a game-theoretic framework to construct highly parallelizable blocks, with penalties imposed for creating overly sequential ones. However, a detailed analysis of these frameworks is beyond the scope of this work and is left for future research.

Therefore, while \sys extends the capabilities of correct nodes to improve the system throughput and empowers them to prevent clients from bottlenecking the system, it does not alter the role a malicious validator could play in the system compared to the state of the art. In fact, due to its modular nature, individual validators on many blockchains could already plug \sys into their stack without requiring a hard fork.


\subsection{Censorship Resistance}

A common concern for leader-based protocols is censorship resistance. In \sys, the leader has, as part of the protocol, the power to delay some transactions to speed up the overall system. 
However, as the leaders in existing protocols such as Ethereum or Aptos already have this power as there is no mechanism that controls this, \sys would not hand the leader stronger censorship powers compared to the state of the art. 

Nonetheless, protocols focused on short-term censorship resistance such as \cite{xue2023bigdipper} might not work out of the box with \sys.
Thus, adjustments to the protocol would be necessary, to only allow a leader to delay a given transaction up to some bounds.
This presents a direct trade-off between performance and short-term censorship resistance. 

Furthermore, protocols focused on fair ordering, such as \cite{8526804} often require the transaction and metadata to be encrypted which strips \sys of the capability to use transaction meta-data to construct ``Good Blocks''. Nonetheless, these approaches are also generally incompatible with hint-based execution schemes as used in \basesys, Sui, or Solana.

\subsection{Transaction Fees \& Client Incentives}

While a malicious leader can arbitrarily delay a client transaction, in \sys correct nodes might also delay client transactions to improve the overall system throughput. However, in some cases, a client might want their transaction to be included with higher urgency even if it accesses very hot resources, e.g. when a bidding process is approaching the time limit.

Integrating a mechanism with \sys that allows client transactions to be included with higher priority is fairly straightforward. Blockchains such as Bitcoin and Ethereum~\cite{bitcoin,ethereum} already use pricing mechanisms to prioritize transaction inclusion. Therefore, a transaction with a higher fee could be transferred to the beginning of the first batch in the batch handler to guarantee its inclusion in the next block.
In fact, similar to the local fee markets in Solana~\cite{solana}, this kind of pricing scheme, in combination with \sys would naturally result in a higher price for accessing hot resources, incentivizing smart contract developers to design their smart contracts with concurrency in mind and incentivizing users to avoid hot resources during system congestion times. This can help to balance the system beyond the already existing throughput advantages of \sys.

%% file: 7concl.tex
\section{Conclusion}
\label{sec:conclusion}

In this work, we presented \sys, a framework, and algorithm to construct highly parallelizable blocks in the context of parallel smart contract execution.
We evaluated \sys extensively under a series of realistic workloads, demonstrating a throughput improvement of up to \sysmax.   Furthermore, in most workloads, this approach leads to lower latency for the majority of transactions, while only delaying those that access hot resources and cause bottlenecks. 
Moreover, \sys not only improves the throughput of the underlying execution engine but also protects blockchains from being bottlenecked by popular applications.
Finally, \sys is highly modular and can be easily integrated into any production blockchain without any security tradeoffs.